\documentclass[conference, 10pt]{IEEEtran}
\IEEEoverridecommandlockouts
\usepackage{cite}
\usepackage{amsmath,amssymb,amsfonts}
\usepackage{algorithmic}
\usepackage{graphicx}
\usepackage{textcomp}
\usepackage{enumerate}  
\usepackage{xcolor}
\usepackage{placeins}
\usepackage{pgfplots}
\usepackage{subcaption}
\usepackage{float}
\usepackage{tikz}

\usetikzlibrary{arrows}

\begin{document}


\title{Inferring Graph Signal Translations as\\Invariant Transformations for Classification Tasks
}

\author{
    \IEEEauthorblockN{Raphaël Baena, Lucas Drumetz, Vincent Gripon}
     \IEEEauthorblockN{ IMT Atlantique and Lab-STICC, \\{name.surname}@imt-atlantique.fr}
}

\maketitle

\begin{abstract}
The field of Graph Signal Processing (GSP) has proposed tools to generalize harmonic analysis to complex domains represented through graphs. Among these tools are translations, which are required to define many others. Most works propose to define translations using solely the graph structure (i.e. edges). Such a problem is ill-posed in general as a graph conveys information about neighborhood but not about directions. In this paper, we propose to infer translations as edge-constrained operations that make a supervised classification problem invariant using a deep learning framework. As such, our methodology uses both the graph structure and labeled signals to infer translations. We perform experiments with regular 2D images and abstract hyperlink networks to show the effectiveness of the proposed methodology in inferring meaningful translations for signals supported on graphs.
\end{abstract}


\section{Introduction}

Translations are among the most fundamental transformations in signal processing. They are often used as a basic building block to define convolutions, Fourier transform, filters and related tools. In machine learning, they can be exploited to define ad-hoc operators that benefit from the underlying simple regular structure of processed signals, such as in the case of Convolutional Neural Networks (CNNs). As a matter of fact, CNNs were introduced because they can be made invariant to translations when combined with downsampling operators, which is often desirable in practice.

Recently, the field of Graph Signal Processing (GSP) arose with the aim of generalizing classical harmonic analysis to irregular domains described using graphs\cite{GSP}. Among the numerous tools that were introduced in this field, translation has attracted a lot of attention~\cite{emergingGSP}. Contrary to the case of $n$D structures, defining translations for graph signals can be challenging. Incidentally, graphs can represent very regular structures (e.g. sensor network) as well as abstract ones (e.g. social network) and the definition of translations and hence harmonic operators should be sensible for these domains.

In the early days of GSP, the Graph Fourier Transform (GFT) was introduced without relying on translations~\cite{discreetsignal}. Convolutions could then be defined by simple pointwise multiplications in the graph spectral domain. And translations were then obtained by particularizing convolutions with Dirac signals. Later in~\cite{Girault} the authors pointed out that this operator was not an isometry. They proposed alternative definitions based on complex exponentials of the Laplacian matrix of the considered graph. Problematically, these operators do not generalize well classical circular translations on signals defined on grid graphs. Using a completely different approach, the authors in~\cite{translationgraph} defined translations of graph signals directly in the vertex domain (without using the GFT), thus providing an actual generalization of classical tools. Still, this approach comes with a large computational complexity, and struggles with abstract and irregular graph structures.

There are fundamental reasons why it is so challenging to define translations for graph signals. One of them is that a graph typically encompasses a notion of neighborhood (or similarity) between its vertices. On the other hand, translations are defined using directions, which are typically not explicitly available or even meaningful when considering a graph~\cite{translationgraph}. In this work, we would like to propose inferring graph signal translations using not only the graph, but also additional information such as annotated signals on this graph.

Our solution builds upon the idea of translational invariance of classification tasks. In more details, given a graph and samples that belong to distinct classes, we aim at inferring operators constrained by the graph structure and that allow to define weight-sharing deep learning architectures that reach high accuracy on the considered classification task. As such, the inferred operators can be interpreted as transformations that are invariant for the considered task. In the case of regular $n$D signals, we would expect these transformations to include classical translations, but also possibly other operators such as directional dilations or contractions. Interestingly, this approach does not require strong assumptions about the regularity of the graph structure, and can thus be deployed even for abstract domains such as relational networks.




\section{Related work}

Let us consider a \emph{graph} $G = \langle V, E \rangle$, where $V$ is a finite set of vertices and $E$ is a set of pair of vertices called the edges. Such a graph can be conveniently expressed using its binary \emph{adjacency matrix} $\mathbf{A}$ defined as: \begin{align}\mathbf{A}[i,j] = \left\{ \begin{array}{ll} 1 & \text{if } (i,j) \in E\\0 &\text{otherwise} \end{array}\right..\end{align}

The \emph{degree matrix} of $G$ is defined as: \begin{align}
    \mathbf{D}[i,j] =  \left\{ \begin{array}{ll} \sum_{i'\in V}{\mathbf{A}[i,i']} & \text{if } i = j \\ 0 &\text{otherwise} \end{array}\right..\end{align}
    
In the field of spectral graph theory, it is common to also introduce the (combinatorial) \emph{Laplacian} of the graph as the matrix defined as $\mathbf{L} = \mathbf{D} - \mathbf{A}$.

In this work, we are interested in processing signals on graphs. A \emph{graph signal} is a vector $\mathbf{s} \in \mathbb{R}^V$. Of particular interest are Dirac signals which are simple one-hot vectors.

The field of GSP introduces tools to manipulate signals on graphs. These tools include convolutions, filtering, smoothing, translations\dots The rationale is that such operators are defined by taking into account the graph structure (i.e. the graph edges). In the particular case where the considered graph is an oriented ring graph, the tools defined by the framework of GSP perfectly match the ones defined for 1D signals \cite{discreetsignal}.

This matching does not necessarily hold for more complex graph structures. In particular, considering regular 2D grid graphs, the operators defined using the GSP toolbox typically differ from the traditional 2D corresponding ones~\cite{translationgraph}. Incidentally, defining a graph signal translation operator is challenging, because a graph structure only encompasses information about neighborhood of vertices and not directionality~\cite{translationgraph}.

In the early days of GSP, translations were defined on top of convolutions. As a matter of fact, the authors in~\cite{emergingGSP} propose a definition of GFT of a signal $\mathbf{s}$ by simply projecting $\mathbf{s}$ to a basis where the Laplacian of the graph is diagonal. The inverse GFT can be obtained by projecting backwards to the canonical basis. Then, in~\cite{spectraltranslation1,spectraltranslation2} the authors define convolutions in three steps: first they compute the GFT of considered signals, then they pointwise multiply their spectral coordinates, and finally they perform an inverse GFT on the resulting vector. Graph signal translations can then be obtained by convolving signals with a Dirac. The authors of~\cite{Girault} point out in their paper that these translations are not isometric. They introduce alternative definitions using complex exponentials of the Laplacian matrix. Problematically, the definitions in~\cite{spectraltranslation1},\cite{spectraltranslation2},\cite{Girault} do not properly generalize translations for signals on graphs, because, as we pointed out previously, these operators typically do not match the expected ones when considering regular 2D grid graphs. As a matter of fact, the translations defined in~\cite{discreetsignal} are isotropic.

In~\cite{motifnet}, the authors aim at identifying directions or relevant graph motifs in order to define graph signal convolutions. These motifs represent meaningful connectivity patterns, e.g triangle motifs which are crucial for social networks~\cite{Benson}. Once a set of motifs is chosen, nonisotropic Laplacians are defined for each one. Convolutions are then defined as multivariate polynomial of the Laplacian matrices. Two key issues with this methods are the huge amount of parameters it relies upon and the difficulty of choosing relevant motifs.

With the purpose of proposing graph signal operators that fully match the expected ones for regular grid graphs, the authors in~\cite{translationgraph} introduce a definition of translations directly in the vertex domain (i.e. that does not use the GFT). In their work they characterize translations as functions $\phi$, defined from a subset of vertices $V'$, that are i) injective ($\phi(v) = \phi(v') \Rightarrow v = v', \forall v,v' \in V'$), ii) edge-constrained ($(v, \phi(v))\in E, \forall v \in V'$) and iii) neighborhood-preserving ($(v,v')\in E \Leftrightarrow (\phi(v),\phi(v')) \in E, \forall v,v' \in V'$). Injectivity and neighborhood-preservation are key characteristics to ensure the matching with regular translations, but they are poorly suited for abstract graph structures such as social networks.

In~\cite{JC}, the authors introduce pseudo-convolutions for deep neural networks that can be seen as implementing the edge constraint previously introduced. Namely, they introduce a tensor $\mathbf{S}$ and a vector $\mathbf{w}$. The binary tensor $\mathbf{S}$ is of dimension $N \times N \times K$, where $N$ is the number of vertices in the considered graph and $K$ is a hyperparameter. Moreover, $\mathbf{S}[i,j,k]$ is zero if $(i,j)\not\in E$, and $\mathbf{S}[i,j,:]$ contains at most one nonzero entry. The vector $\mathbf{w}$ contains $K$ coordinates. The tensor-matrix product along the third mode of $\mathbf{S}$ by $\mathbf{w}$, denoted as  $\mathbf{S}$ $\times_3$ $\mathbf{w}$ creates a $N\times N$ matrix $\mathbf{W}$  that can be seen as a weighted version of the adjacency matrix $A$ of the considered graph. The authors show that for particular choices of $\mathbf{S}$, they can retrieve classical convolutions for regular grid graphs. More generally, slices $\mathbf{S}[:,:,k]$ can be interpreted as graph signal translations. In this paper, we propose to infer the tensor $\mathbf{S}$ using both the graph structure and a set of labeled signals.

\section{Problem Statement and Methodology}

The rationale behind CNNs is to exploit the invariance of input labels to translations~\cite{LeCun}, which is achieved through weight sharing schemes. In more details, translations are used to define convolutions. When convolutions are combined with pooling operations, they can produce representations that are invariant with respect to translations. Resulting CNNs can obtain significant gains in accuracy compared to translation-agnostic architectures such as multi-layer perceptrons~\cite{LeCun}. 

The key idea of our proposed methodology is to reverse this reasoning. Namely, we propose to define learnable operators that are aligned with the graph structure, from which we build pseudo-convolutions by learning ad-hoc weight sharing schemes. Combined with pooling, we obtain architectures that can be trained end-to-end to solve classification tasks. Once a network with good performance is found, we can then assimilate our learned operators as pseudo-translations, or more generally classification invariant operations.



In more details, let us consider a simple example where the graph is a ring with adjacency matrix $\mathbf{A}$. Let us suppose that we consider a periodic graph signal $\mathbf{s}$ made of $N=4$ dimensions, on which we can operate $k=3$ translations denoted through their matrix representations $\left(\mathbf{T}_k\right)$, where $\mathbf{T}_k \in\mathbb{R}^{N \times N}$. For this simple example $T_0=\begin{pmatrix}
1 & 0 & 0& 0\\
0 & 1 & 0& 0\\
0 & 0 & 1& 0\\
0 & 0 & 0& 1\\
\end{pmatrix}$ will be the identity, $T_1=\begin{pmatrix}
0 & 0 & 0& 1\\
1 & 0 & 0& 0\\
0 & 1 & 0& 0\\
0 & 0 & 1& 0\\
\end{pmatrix}$ and $T_2=\begin{pmatrix}
0 & 1 & 0& 0\\
0 & 0 & 1& 0\\
0 & 0 & 0& 1\\
1 & 0 & 0& 0\\
\end{pmatrix}$ circular translations corresponding to the two orientations of the ring. We build a tensor $\mathbf{S}\in \mathbb{R}^{N\times N\times K}$ by concatenating matrices $\left(\mathbf{T}_k\right)_k$. We also define a convolutional kernel vector $\mathbf{w}$ indexed by the $K$ possible translations.
Then it holds that: \begin{eqnarray} \mathbf{S} \times_3 \mathbf{w} &=& \sum_k \mathbf{\mathbf{w}}[k]\mathbf{S}[:,:,k]\ \nonumber \\ 
&=& \sum_k \mathbf{w}[k] (\mathbf{T}_k) \nonumber\\
&=&\begin{pmatrix}
w_0 & w_2 & 0& w_1\\
w_1 & w_0 & w_2& 0\\
0 & w_1 & w_0& w_2\\
w_2 & 0 & w_1& w_0\\
\end{pmatrix}.
\label{Eqconv}\end{eqnarray}

We indeed recognize a Toeplitz circulant convolution matrix. These equations can be generalized for any regular $n$D graph easily, and to any graph by constraining the structure of $\mathbf{S}$. The graph convolution operation $\star$ can be then defined as:
\begin{align}
    \mathbf{s} \star \mathbf{w} = \mathbf{s}^{\top}(\mathbf{S}\times_3 \mathbf{w}).
\end{align}

We propose to learn matrices $(\mathbf{T}_k)_k$ by optimizing a deep neural network meant to classify graph signals, under the constraint that $\mathbf{T}_k[i,j] \neq 0 \Leftarrow \mathbf{A}[i,j] \neq 0$, where $\mathbf{A}$ is the graph adjacency matrix. In other words, $(\mathbf{T}_k)_k$ are edge constrained transformations.

\subsection{Problem Statement}

\tikzstyle{int}=[draw, fill=blue!20, minimum size=2em]
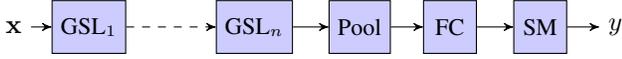
\begin{figure}
\centering
\begin{tikzpicture}[node distance=1.5cm,auto]
    \node(a) at (0,0) {$\mathbf{x}$};
    \node(b)[int] at (1,0) {\small{GSL}$_1$};
    \node(c)[int] at (3.2,0) {\small{GSL}$_n$};
    \node(d)[int] at (4.6,0) {\small{Pool}};
    \node(e)[int] at (5.8,0) {\small{FC}};
    \node(f)[int] at (7,0) {\small{SM}};
    \node(g) at (8,0) {$y$};
    \path[->,>=stealth']
    (a) edge (b)
    (c) edge (d)
    (d) edge (e)
    (e) edge (f)
    (f) edge (g)
    ;
    \path[->,>=stealth',dashed]
    (b) edge (c);
\end{tikzpicture}

\caption{Depiction of the used deep learning architecture. GSL stands for Graph-Signal Layer, Pool for a global average pooling, FC for a fully connected layer and SM for a \texttt{softmax}.}
\label{archicture}
\end{figure}

For the sake of simplicity, we describe here the processing of tensors with only one filter, that is to say a single $\mathbf{w}$. Note that all the equations of this section could be generalized to the case of multiple filters, which boils down to adding a dimension to all tensors and computations presented thereafter.

Let us recall that a deep neural network can be described by a function $f$ mapping the input to the output. The function $f$ is obtained by assembling elementary functions, called layers, that are most of the time of the form: $\mathbf{x} \mapsto \sigma(\mathbf{W} \mathbf{x} + \mathbf{b})$, where $\mathbf{W}$ is a weight matrix, $\mathbf{b}$ is a bias vector and $\sigma$ is a nonlinear function, usually parameter-free and applied component-wise. The weight matrix and its associated bias vector are the trainable parameters $\theta$ of the network.

In the case of classification, the aim is to train $f$ to map raw inputs (e.g. images) to their corresponding class. For that matter, we typically use two datasets, a training one, denoted $\mathcal{D}_{train}$, that is used to learn the parameters and a validation one used to stress the ability of the trained function $f$ to correctly predict the class of previously unseen inputs. Also, the network function $f$ ends by applying a $\texttt{softmax}$ operator.

The most typical setting for training a classifier is to rely on a cross-entropy loss function $\mathcal{L}$. Denoting $(\mathbf{x}, y) \in \mathcal{D}_{train}$ where $\mathbf{x}$ is an input and $y$ its corresponding output, we have: $$\mathcal{L}(\mathbf{x},y) = - \log(f(\mathbf{x})[y]).$$

The deep neural network function $f$ is optimized to solve the following problem:
$$\arg\min_{\theta}{\sum_{(\mathbf{x},y)\in \mathcal{D}_{train}}{\mathcal{L}(\mathbf{x},{y})}}.$$ 

In practice, variants of the Stochastic Gradient Descent algorithm are often used for this optimization.

Of particular interest for vision tasks are convolutional layers, in which the weight tensor $\mathbf{W}$ implements a convolution operator. In our case, we do not have an explicit access to translations, hence to convolutions. Thus we rather make use of Graph-Signal Layers (GSLs): $\mathbf{s} \mapsto \sigma(\mathbf{s}^\top (\mathbf{S} \times_3 \mathbf{w}))$, where slices $\mathbf{S}[:,:,k]$ are edge-constrained: $$\mathbf{S}[i,j,k] \neq 0 \Leftarrow \mathbf{A}[i,j] \neq 0.$$

Given Equation~\eqref{Eqconv}, this layer is a generalization of convolutional layers. Multiple GSLs can be defined, each with its own weight vector $\mathbf{w}$, but sharing the same global $\mathbf{S}$.

Let us now imagine that we are a given a deep neural network function $f$, with parameters $ \{\mathbf{S}, \omega,\theta \}$, containing some GSLs.  Here $\mathbf{S}$ represents the graph transformations (which are implicit in CNNs), $\omega$ are the parameters of the GSLs, and $\theta$ are the remaining parameters (e.g. for fully connected layers).  The problem we aim at solving is to find: $$\arg\min_{\mathbf{S}, \omega,\theta}{\sum_{(\mathbf{x},y)\in \mathcal{D}_{train}}{\mathcal{L}(\mathbf{x},y)}}.$$ 

Specifically, we are interested in solutions in which $\mathbf{S}[i,:,k]$ are one-hot vectors, so that slices $\mathbf{S}[:,:,k]$ can be interpreted as pseudo-translations. In the next subsection, we delve in more details in how we propose to enforce this constraint.

\subsection{Methodology}
\begin{figure*}[!ht]
\centering
\begin{subfigure}{.18\textwidth}
  \centering
  \includegraphics[width=1\linewidth]{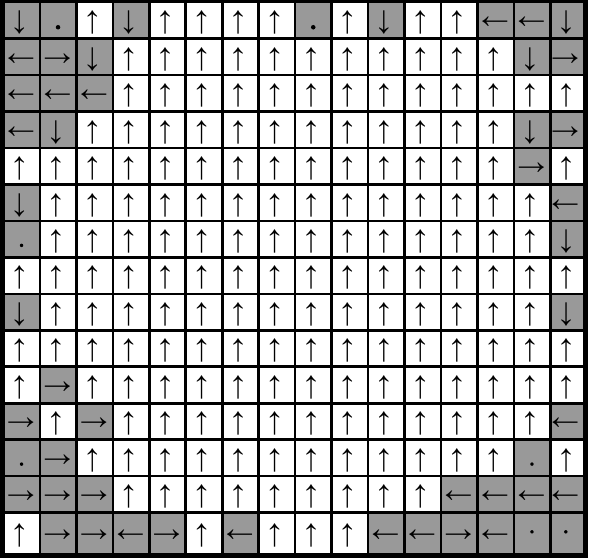}
  \caption{$T_0$}
  \label{fig:T0}
\end{subfigure} \hspace{1mm} 
\begin{subfigure}{.18\textwidth}
  \centering
  \includegraphics[width=1\linewidth]{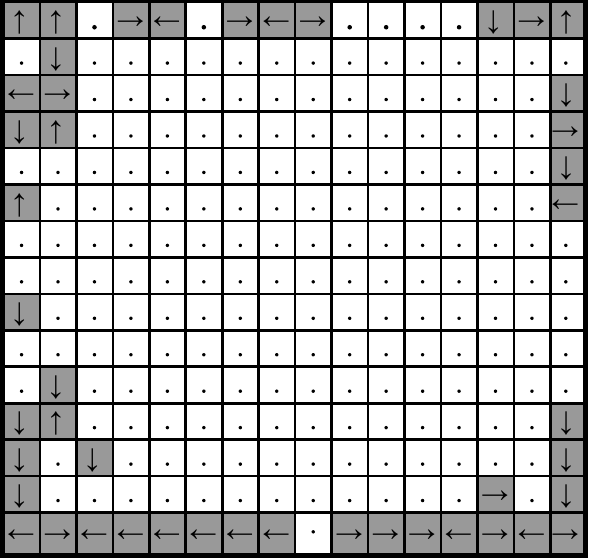}
  \caption{$T_1$ }
  \label{fig:T1}
\end{subfigure}
\begin{subfigure}{.18\textwidth}
  \centering
  \includegraphics[width=1\linewidth]{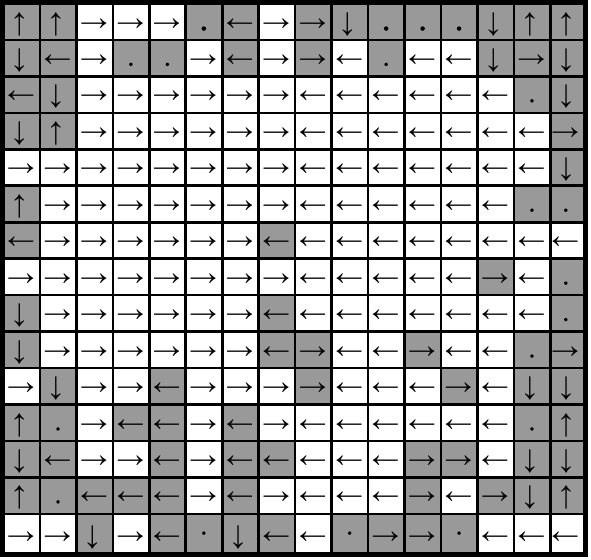}
  \caption{$T_2$ }
  \label{fig:T2}
\end{subfigure}
\begin{subfigure}{.18\textwidth}
  \centering
  \includegraphics[width=1\linewidth]{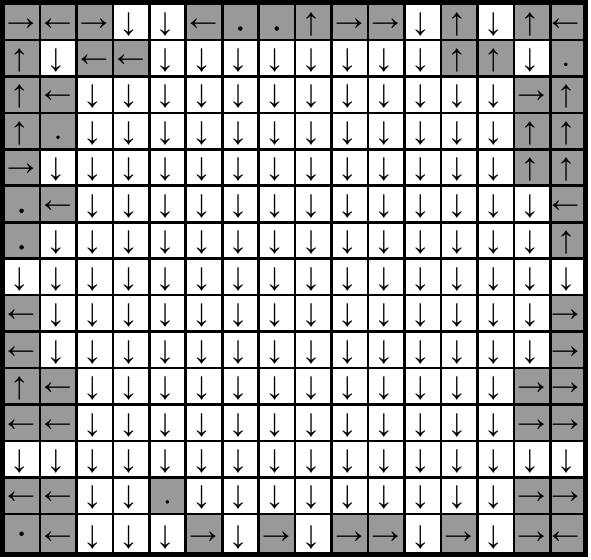}
  \caption{$T_3$}
  \label{fig:T3}
\end{subfigure}
\begin{subfigure}{.18\textwidth}
  \centering
  \includegraphics[width=1\linewidth]{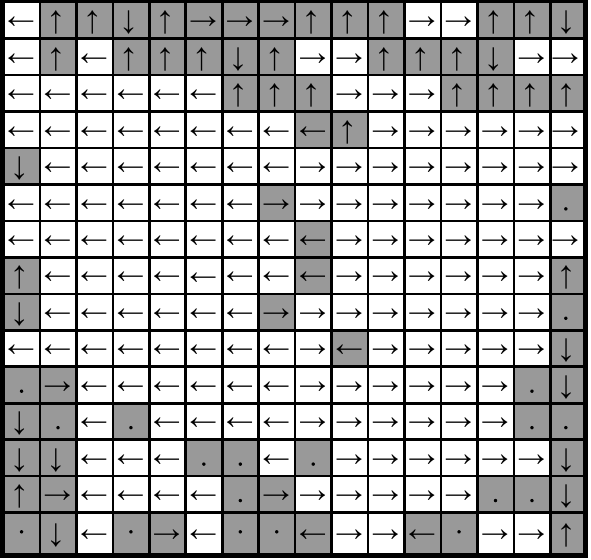}
  \caption{$T_4$}
  \label{fig:T4}
\end{subfigure}
\caption{Depiction of inferred pseudo-translations when considering the CIFAR-10 dataset on a regular 2D grid graph.}
\label{fig:regulartranslation}
\end{figure*}

As stated in the introduction, convolutional neural networks, when they are built with pooling layers, have the asset of producing translation invariant decisions. However, performing pooling on graph signals can be hard, because it requires computing graph downsampling~\cite{emergingGSP}. This is why in this paper, we adopt a simple workaround where we only perform a single pooling operation at the penultimate layer of our proposed architecture, right before the final fully connected layer. This pooling is global, so that it completely shrinks the graph dimension: all vertices values are averaged in a single value, for each considered filter. A depiction of the proposed architecture is available in Figure~\ref{archicture}.

Optimizing deep neural network functions over a discrete domain is a hard task~\cite{courbariaux}, since it involves binary matrix constraints, which are not straightforward to enforce. Because our aim is to obtain one-hot vectors, which is similar to~\cite{Ghouthi}, we adopt the same strategy. Namely, we apply a $\texttt{softmax}$ operator over the second dimension of $\mathbf{S}$, with a varying temperature $t$ ($\mathbf{x} \mapsto \texttt{softmax}(\mathbf{x}/t)$). This temperature starts with value $t_{init}$, typically large, in which case the $\texttt{softmax}$ operator has the effect of making the lines $\mathbf{S}[i,:,k]$ constant where defined (recall that $\mathbf{S}[:,:,k]$ is edge-constrained). At the end of the training, the final temperature is $t_{final}$, typically small, so that the $\texttt{softmax}$ boils down to a regular $\max$ operator, transforming lines $\mathbf{S}[i,:,k]$ into one-hot vectors.

We experimented with various strategies to interpolate the temperature between $t_{init}$ and $t_{final}$. Our most consistent results were obtained using an exponential interpolation: $$t(s) = t_{init} \left(t_{final}/t_{init}\right)^{s/s_{total}},$$ where $s$ is the current step in the training phase, and $s_{total}$ is the total number of steps used for training. At the end of the training process, we use a temperature of 0 to interpret the slices of~$\mathbf{S}[:,:,k]$ as pseudo-translations. In the next section, we present experiments on toy and real datasets.

\section{Experiments}

In this section we present experiments on various types of graphs from very regular structures (images supported on 2D grid graphs) to abstract ones (hyperlink networks). We evaluate our method on two datasets: CIFAR-10~\cite{Cifar10} and webKB~\cite{webKB}. CIFAR-10 is a classification dataset of images made of 32$\times$32 pixels with three primary colors grouped in 10 classes. 
WebKB is a dataset composed of 877 web pages from computer science departments of universities  classified into one of five classes (student, project,
course, staff, and faculty). The dataset contains word-based feature vectors of dimension 1703 for each of the websites, as well as a hyperlink graph. This dataset is typically used in contexts of semi-supervised classification, where only a portion of the websites are labeled.



\begin{figure}[!h]
\centering
\begin{subfigure}{.15\textwidth}
  \centering
  \includegraphics[width=1\linewidth]{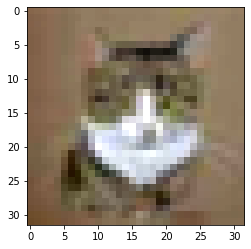}
  \caption{original image }
  \label{fig:originalimage}
\end{subfigure}
\begin{subfigure}{.15\textwidth}
  \centering
  \includegraphics[width=1\linewidth]{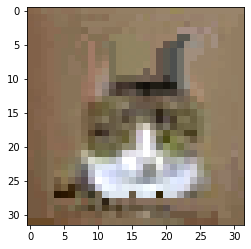}
  \caption{$T_0$}
  \label{fig:T0}
\end{subfigure}
\begin{subfigure}{.15\textwidth}
  \centering
  \includegraphics[width=1\linewidth]{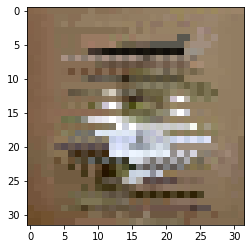}
  \caption{$T_1$ }
  \label{fig:T1}
\end{subfigure}
\begin{subfigure}{.15\textwidth}
  \centering
  \includegraphics[width=1\linewidth]{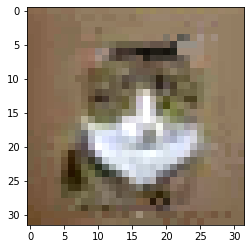}
  \caption{$T_2$ }
  \label{fig:T2}
\end{subfigure}
\begin{subfigure}{.15\textwidth}
  \centering
  \includegraphics[width=1\linewidth]{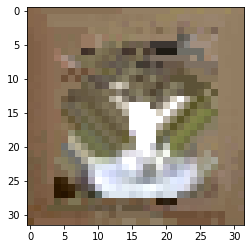}
  \caption{$T_3$ }
  \label{fig:T3}
\end{subfigure}
\begin{subfigure}{.15\textwidth}
  \centering
  \includegraphics[width=1\linewidth]{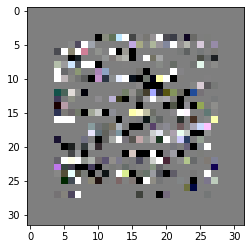}
  \caption{translation of\cite{spectraltranslation1} }
  \label{fig:T4}
\end{subfigure}
\caption{Inferred translations $T_0,T_1,\cdots,T_3$ and comparison with the translation defined in \cite{spectraltranslation1} on a near-regular graph.}
\label{fig:irregulartranslation}
\end{figure}

\subsection{Sanity check with regular grid graphs}

In our first experiment, we aim at verifying the ability of our proposed method to retrieve classical translations when dealing with 2D signals and structures. To this end, we use the CIFAR-10 dataset downscaled to 16$\times$16 pixel images and suppose given a regular grid graph for supporting the image signals. In more details, the grid graph is such that a vertex corresponds to a pixel, and each pixel is connected through the edges to its four direct neighbors.

In Figure~\ref{fig:regulartranslation}, we depict the result of our proposed method. An inferred pseudo-translation $\mathbf{T}$ is represented in a grid of size $16\times 16$. For each vertex we represent by an arrow the neighbor vertex it is associated with through $\mathbf{T}$ (recall that inferred pseudo-translations are edge-constrained, so that this representation is well defined). For each $\mathbf{T}_k$, we highlight the vertices that correspond to the majority direction. Interestingly, we observe that $\mathbf{T}_0$ and $\mathbf{T}_3$ tend to approach regular translations. Note that $\mathbf{T}_1$ is almost the identity function. Surprisingly, we observe that $\mathbf{T}_4$ and $\mathbf{T}_2$ resembles respectively an horizontal dilation and compression. As a matter of fact, such transformations are valid in our framework and would typically be invariant for the classification problem at hand.


\subsection{Experiments with a near-regular inferred graph structure}
    
In our second experiment, we use an inferred graph structure that is obtained by computing the covariance matrix from the training set of CIFAR-10, and thresholding to keep only the 5 nearest neighbors of each vertex (including self-loops). The inferred graph structure is not as regular as the previously used 2D grid graph even though it remains similar.

Due to the non-regular structure of the graph it is not possible to use the same representation than in Figure~\ref{fig:regulartranslation}. Therefore we illustrate the obtained transformations by applying them directly on an arbitrarily selected input image. Results are shown in Figure~\ref{fig:irregulartranslation}. We can clearly see that obtained transformations are not exactly classical translations, but most of them are interpretable: $\mathbf{T}_0$ look likes a vertical translation, $\mathbf{T}_2$ the identity, $\mathbf{T}_1$ and $\mathbf{T}_3$  horizontal dilation and contraction.

\subsection{Experiments with hyperlink networks}

To illustrate the genericity of the approach, we next run an experiment with the WebKB dataset. For lack of a better method to evaluate the obtained transformations, we compare the accuracy achieved using the proposed methodology with a standard method from the literature: graph convolutional neural network GCN~\cite{GCNref}. We averaged the obtained accuracy on 10 different splits of training/validation/test sets. GCN obtains an average of $86\%$ and our method $87\%$. Note that GCN and the proposed methodology reach similar performance, yet the two systems are quite different: GCN uses isotropic diffusion of signals, whereas we focus on directional inferred translations. Moreover, contrary to GCN, our approach is not designed to optimize classification performance but to infer meaningful edge-constrained transformations.

\subsection{Influence of hyperparameters}

\begin{figure}
    \centering
\begin{tikzpicture}

\begin{semilogxaxis}[
    title={},
    xlabel={initial temperature $t_{init}$ },
    ylabel={distance, accuracy},
    xmin=0.1, xmax=1000,
    ymin=0, ymax=80,
    xtick={0.1, 1,10, 100,1000 },
    ytick={0,20,40,60,80},
    yticklabels={0\%,20\%,40\%,60\%,80\%},
    legend pos=north west,
    ymajorgrids=true,
    grid style=dashed,
    height=5cm,
       width=.41\textwidth
]

       \addplot[
    color=green,
    mark=square,
    ]
    coordinates {
    (0.1,69)( 1,37.5)(10,77)(50,41)(100,20)(200,40)
    (500,53)(1000,30.8)
    };
    \legend{};

\addplot[
    color=purple,
    mark=square,
    ]
    coordinates {
    (0.1,60)( 1,57)(10,58)(50,46)(100,29)(200,41)
    (500,50)(1000,34)
    };
    \legend{}
\addplot[
    color=cyan,
    mark=square,
    ]
    coordinates {
    (0.1,52)( 1,26.5)(10,22.2)(50,11)(100,23)(200,18.7)
    (500,22)(1000,21)
    };
    \legend{}
\addplot[
    color=orange,
    mark=square,
    ]
    coordinates {
    (0.1,66.6)( 1,58)(10,37)(50,16)(100,21)(200,25.7)
    (500,22)(1000,23.4)
    };
    \legend{}
    array([0.619   , 0.4475  , 0.48525 , 0.2875  , 0.2325  , 0.313   ,
       0.318925, 0.273   ])
    \addplot[
    color=red,
    mark=square,
    ]
    coordinates {
    (0.1,61.9)( 1,44.75)(10,48.5)(50,28.75)(100,23.25)(200,31.3)
    (500,31.8)(1000,27.3)
    };
    \legend{}

    \addplot[
    color=black,
    mark=triangle,
    ]
    coordinates {
    (0.1,60.4)( 1,66.6)(10,64)(50,66)(100,65.3)(200,61.17)(500,61.19)(1000,63.1)
    };
    \legend{}
     \end{semilogxaxis}
\end{tikzpicture}

    \caption{Impact of $t_{init}$ on the accuracy (black) and distance of the obtained translation : identity (orange), up (green), down (purple), dilation (blue), and the average distance (red). }
    \label{fig:tinit}
\end{figure}
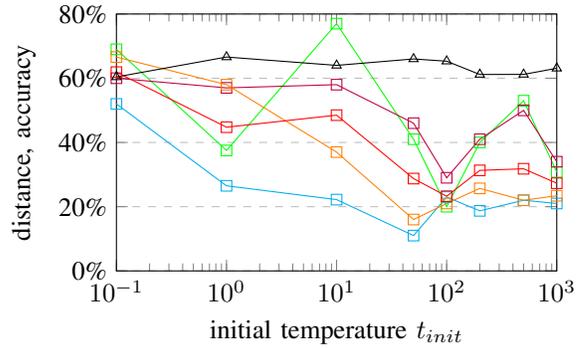

Finally, in a last series of experiments, we illustrate the sensitivity of the proposed method with respect to the hyperparameters $t_{init}$ and $t_{final}$. In Figure~\ref{fig:tfinit}, we fix $t_{init}$ and vary $t_{final}$, whereas in Figure~\ref{fig:tinit}, we fix $t_{final}$ and vary $t_{init}$. In these experiments we evaluate the impacts of the initial and final temperatures on the accuracy of the network and the transformations obtained. The ``distance'' measures the number of differences between obtained transformations and the closest 2D translation, dilation or contraction. For this evaluation we use CIFAR-10 dataset and assume that the images rely on the grid-graph. As can be observed, the method is quite robust to changes in these hyperparameters. 

\begin{figure}
    \centering
\begin{tikzpicture}

\begin{semilogxaxis}[
    title={},
    xlabel={final temperature $t_{final}$ },
    ylabel={distance, accuracy},
    xmin=0.0001, xmax=10,
    ymin=0, ymax=80,
    xtick={0.0001, 0.001, 0.01, 0.1, 1, 10 },
    ytick={0,20,40,60,80},
    yticklabels={0\%,20\%,40\%,60\%,80\%},
    legend pos=north west,
    ymajorgrids=true,
    grid style=dashed,
    height=5cm,
    width=.41\textwidth
]

       \addplot[
    color=green,
    mark=square,
    ]
    coordinates {
    (0.0001,41)( 0.001,24)(0.01,26)(1,68)(10,69)
    };
    \legend{};

\addplot[
    color=brown,
    mark=square,
    ]
    coordinates {
    (0.0001,40)( 0.001,21)(0.01,38)(1,69)(10,59)
    };
    \legend{}
\addplot[
    color=cyan,
    mark=square,
    ]
    coordinates {
    (0.0001,22)( 0.001,20)(0.01,22)(1,26)(10,42)
    };
    \legend{}
\addplot[
    color=orange,
    mark=square,
    ]
    coordinates {
    (0.0001,27)( 0.001,15)(0.01,29)(1,44)(10,40)
    };
    \legend{}
    \addplot[
    color=red,
    mark=square,
    ]
    coordinates {
    (0.0001,32.5)( 0.001,20)(0.01,31.3)(1,51.8)(10,52.4)
    };
    \legend{}

    \addplot[
    color=black,
    mark=triangle,
    ]
    coordinates {
    (0.0001,60)( 0.001,64.71)(0.01,66)(1,56)(10,53)
    };
    \legend{}
     \end{semilogxaxis}
\end{tikzpicture}
    \caption{Impact of $t_{finit}$ on the accuracy (black) and distance of the obtained translation : identity (orange), up (green), down (purple), dilation (blue), and the average distance (red). }
    \label{fig:tfinit}
\end{figure}

\section{Conclusion}
In this paper we have introduced a new methodology based on deep learning to infer graph signal translations from both a graph structure and a set of labeled signals. We empirically showed that this methodology is able to retrieve usual 2D translations from regular images. We also conducted experiments on an abstract hyperlink network and obtained performance similar to that of state-of-the-art. There are many open questions following this work, including other possible ways to infer translations using labeled graph signals, better choice of hyperparameters, design of deep learning architectures and of the classification dataset.

\bibliographystyle{IEEEtran}
\bibliography{biblio}

\end{document}